\documentclass[twocolumn,aps,prl,groupedaddress]{revtex4-1}%
\usepackage{graphicx}
\usepackage{dcolumn}
\usepackage{bm}
\usepackage{color}
\usepackage{dcolumn}
\usepackage{amsmath}
\usepackage{multirow}
\usepackage{amsfonts}
\usepackage{amssymb}
\usepackage{lineno}
\usepackage[colorinlistoftodos]{todonotes}%
\usepackage[colorlinks, linkcolor=blue, citecolor=blue, urlcolor=blue]{hyper
ref}%

\providecommand{\U}[1]{\protect\rule{.1in}{.1in}}
\fontfamily{cmss}
\providecommand{\U}[1]{\protect\rule{.1in}{.1in}}

\begin{document}
\title{Magnetism-driven
unconventional effects in Ising superconductors: role of proximity, tunneling, and nematicity}
\author{Darshana Wickramaratne}
\affiliation{Center for Computational Materials Science, U.S. Naval Research Laboratory,
Washington, DC 20375, USA}
\email{darshana.wickramaratne@nrl.navy.mil}
\author{Menashe Haim}
\affiliation{The Racah Institute of Physics, The Hebrew University of Jerusalem, Jerusalem
9190401, Israel}
\author{Maxim Khodas}
\affiliation{The Racah Institute of Physics, The Hebrew University of Jerusalem, Jerusalem
9190401, Israel}
\author{I.I. Mazin}
\affiliation{Department of Physics and Astronomy, George Mason University, Fairfax, VA
22030, USA}
\affiliation{Quantum Science and Engineering Center, George Mason University, Fairfax, VA
22030, USA}
\date{\today}

\begin{abstract}
Hybrid Ising superconductor-ferromagnetic insulator heterostructures provide a
unique opportunity to explore the interplay between proximity-induced
magnetism, spin-orbit coupling and superconductivity.
Here
we use a combination of first-principles calculations of NbSe$_{2}$/CrBr$_{3}$
heterostructures and an analytical theory of Ising superconductivity 
to analyze the existing experiments and provide a complete explanation of highly nontrivial and largely
counterintuitive effects:  an {\it increase} in the magnitude of the
superconducting gap accompanied by the {\it broadening} of the tunneling peaks; hysteretic
behavior of the tunneling conductance that sets in $\approx 2$ K below $T_c$; and nematic symmetry breaking
in the superconducting state. 
The microscopic reason in all three cases appears to be the interplay
between the proximity-induced exchange splitting and intrinsic defects.
Finally, we predict additional interesting effects that at the moment cannot
be addressed experimentally: spin-filtering when tunneling across CrBr$_{3}$ 
and tunneling ``hot spots'' in momentum space that are anticorrelated with regions
where the spin-orbit splitting is maximum.
\end{abstract}
\maketitle

One of the most intriguing discoveries in superconductivity in the last decade is the 
so-called Ising superconductivity, which appears in materials without inversion symmetry 
and with a particular type of spin-orbit coupling (SOC) 
\cite{lu2015evidence,xi2016ising,sergio2018tuning,ising_prx,shaffer2020crystalline,tang2021magnetic}. Thus far all experimental work on 
Ising superconductivity have been performed on single layers of the transition metal dichalcogenides (TMD), such as NbSe$_2$. 
Unlike conventional superconductors which can be classified by parity (centrosymmetric materials) or
by the leading parity (non-centrosymmetric materials), Ising superconductivity represents a qualitatively different class, where
each Cooper pair is described by an equal mix of singlet and triplet wave functions \cite{ising_prx,shaffer2020crystalline}.
This manifests in a range of unique properties which includes a theoretically infinite thermodynamical
critical field along certain directions and nontrivial interplay of superconductivity with magnetism.

Combined with developments in the field of two-dimensional magnetic semiconductors \cite{mcguire2017crystal,kim2019evolution}
this has motivated a large effort focused on using Ising superconductors in 2D
Josephson junctions \cite{dvir2018spectroscopy}, or investigating tunneling across 
magnetic tunnel barriers \cite{tokuyasu1988proximity,tedrow1986spin,heikkila2019thermal}.
Superconductor/ferromagnetic insulator junctions have, in particular, been used to elucidate the
fundamental properties of the superconducting contacts and are also pursued 
for applications in spintronics \cite{eschrig2015spin} or hosting
topological states
\cite{sau2012experimental,glodzik2020engineering,kezilebieke2020topological}. 

Recent experiments \cite{hamill2021two,cho2020distinct,
kim2019tailored,idzuchi2020van,kang2021giant,ai2101van} indicate that the behavior of
Ising superconductor-magnetic insulator junctions is qualitatively different
compared to conventional superconductors.
Some of the most puzzling observations include hysteretic behavior 
in NbSe$_{2}$/Cr$_2$Ge$_2$Te$_6$ \cite{idzuchi2020van,ai2101van}
and in NbSe$_{2}$/CrBr$_{3}$/NbSe$_{2}$ 
 heterostructures \cite{kang2021giant}, which only appeared at $\sim$ 2K below $T_c$.
Kang {\it et al} \cite{kang2021giant} convincingly demonstrated that the hysteresis, inexplicably, emerges from the Ising superconductor, 
and not from the ferromagnetic insulator.  Refs.~\cite{hamill2021two, cho2020distinct} reported evidence of
a two-fold rotation symmetry of the superconducting state, violating 
the three-fold symmetry of the hexagonal lattice of NbSe$_{2}$. 
Finally, Ref.~\cite{kang2021giant} also found that as an external
in-plane magnetic field rotates the CrBr$_{3}$ spins from being along 
$\hat{z}$ to being in-plane, the superconducting gap,
$\Delta$, increases by $\sim$2\%, while the broadening of the tunneling peak at
the same time \textit{also increases} by $\sim$50\%. This is
counterintuitive: one expects that when $\Delta$ increases
the width of the tunneling peaks should decrease.  

These experimental observations contain a lot of interpretative power
and form a three-pronged puzzle that we will provide microscopic insight into
in this study using a combination of first-principles calculations and 
analytical calculations based on a theory of Ising superconductivity
that also accounts for spin-conserving scattering due to paramagnetic
point defects  \cite{nguyen2017atomic,wang2017high,sosenko2017unconventional,mockli_JAP20,Haim2020}.

\begin{figure}[!th]
\includegraphics[width=8.5cm]{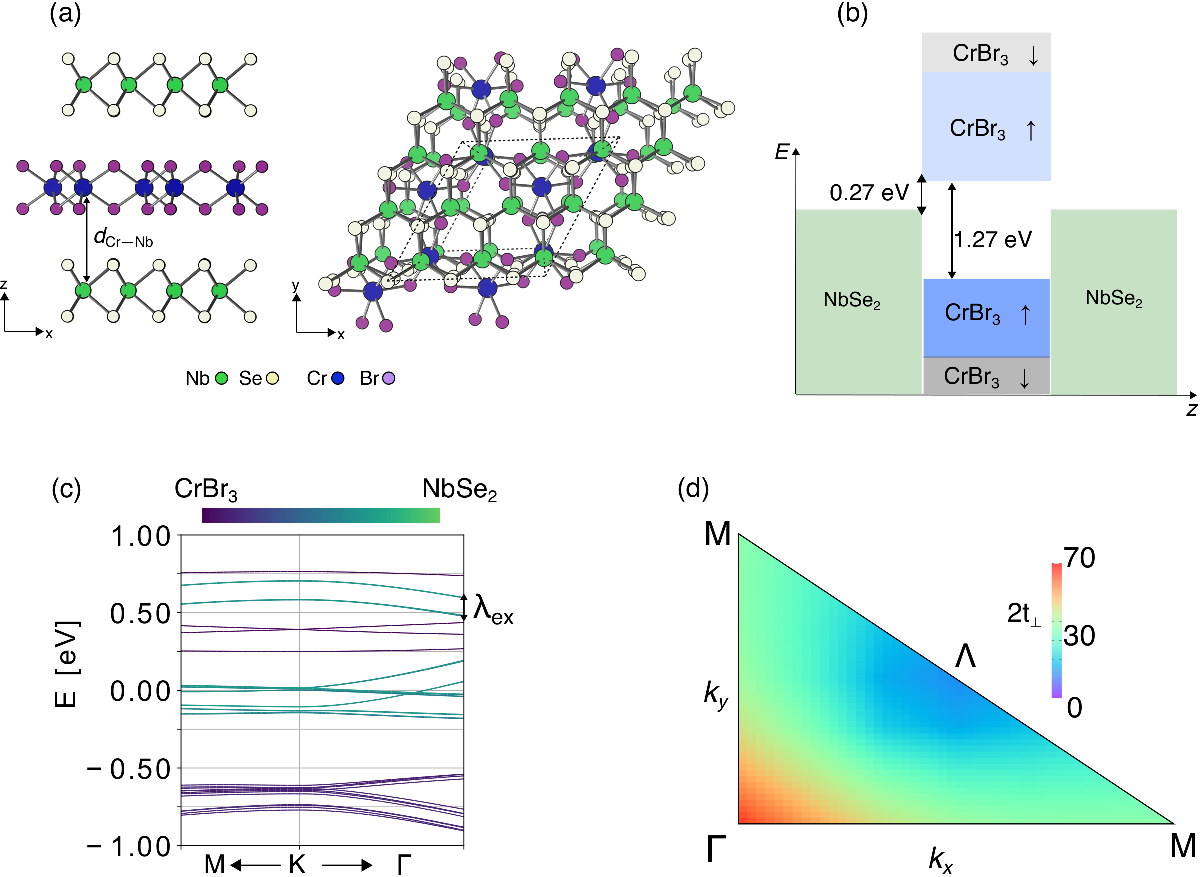}\caption{
(a) Trilayer heterostructure showing the
side view and the top view. 
(b) Alignment of the
energy levels of NbSe$_{2}$ with respect to monolayer CrBr$_{3}$ at the
\textrm{K}-point. (c) Spin-polarized band structure of the trilayer
heterostructure around the \textrm{K} point. 
(d)  Interlayer coupling, 2t$_{\perp}$, of the 
NbSe$_{2}$/CrBr$_{3}$/NbSe$_{2}$ trilayer heterostructure as a function of 
momentum.  $\Lambda$ corresponds to the $\Gamma$-{\rm K} midpoint.  
}%
\label{fig:struct}%
\end{figure}

We begin by examining the electronic structure of the
NbSe$_{2}$/CrBr$_{3}$/NbSe$_{2}$ trilayer heterostructure using
first-principles calculations \cite{SM}.
The atomic structure of the heterostructure with the lowest energy
is illustrated in Fig.~\ref{fig:struct}(a).
The alignment of the NbSe$_2$ states at the Fermi level 
at {\rm K} with respect to the CrBr$_{3}$ spin
up and spin down states are illustrated in
Fig.~\ref{fig:struct}(b) and the spin-polarized band structure
of the trilayer heterostructure is shown in Fig.~\ref{fig:struct}(c).
The NbSe$_{2}$ states reside
within the spin-up gap of CrBr$_{3}$, close to the spin up conduction band
states of CrBr$_{3}$.

One striking change in the electronic structure of the heterostructure is the
large exchange splitting, $\lambda_{\mathrm{ex}}$, of the NbSe$_{2}$ derived
states (the bias field $\mu_{B}B=\lambda_{ex}/2)$. For the
heterostructure in Fig.~\ref{fig:struct}(a), $\lambda_{\mathrm{ex}} $
is 121 meV between the spin up and spin down states. 
In bilayer NbSe$_{2}$, the two pairs of 
spin degenerate bands, contributed by each monolayer, 
are split due to interlayer coupling, $t_{\perp}$ \cite{SM}. In the
heterostructure calculations, the Nb atoms in the top and bottom
monolayers aquire a magnetic moment, $m_{\rm Nb}$, of $\sim$ 0.10 to 0.13
$\mu_{B}$ due to proximity induced Cr-Nb coupling. This
manifests in a proximity induced exchange splitting, $\lambda_{\rm ex}$, illustrated in
Fig.~\ref{fig:struct}(c), that breaks the spin degeneracy of these bands.

The magnitude of $\lambda_{\mathrm{ex}}$ reflects the magnitude of orbital
overlap between the Nb and Cr $d-$electrons. It crucially depends on
vertical
separation distance between the Nb and Cr atoms, $d_{\mathrm{Cr-Nb}} $ \cite{SM}.
Moreover, as Fig. ~\ref{fig:struct}(b) illustrates, the quasiclassical tunneling barrier
in the spin-majority channel is 0.27 eV, while it is several times larger in the spin 
minority channel. Thus, we predict a strong spin-filtering effect for the Cr spins aligned along $z$
with the spin-minority tunneling being strongly suppressed.

To form a commensurate trilayer heterostructure, we assume 
CrBr$_{3}$ layer is under biaxial tensile strain.
In reality since the lattice mismatch is large, the two layers
are incommensurate, which would lead to to spatially varying stacking of NbSe$_{2}$ with respect to
CrBr$_{3}$. Given the strong itinerancy of Nb electrons, the lateral rigidity
of both layers,
the effective overlap and $\lambda_{\mathrm{ex}}$ should be
averaged over all possible mutual orientations between the two layers, while
the equilibrium distance corresponds to the sterically least favorable geometry,
\textit{i.e.}, when Br and Se ions are aligned vertically. 
The effect of this averaging \cite{SM} is
that $\lambda_{\mathrm{ex}}$ at $d_{\mathrm{Cr-Nb}}\approx6.88$\AA ~which,
according to our calculations is the maximal possible separation distance
between NbSe$_{2}$ and CrBr$_{3}$, becomes $\left\langle \lambda_{\mathrm{ex}%
}\right\rangle \approx0.04$ meV, which is equivalent to a magnetic exchange
field, $B\approx$ 0.7 T.

Inserting a single layer of CrBr$_{3}$ increases the interlayer separation
between the NbSe$_{2}$ layers, which changes $t_{\perp}$. 
We illustrate the magnitude of 2$t_{\perp}$
along $\Gamma$-M and along $\Gamma-\Lambda$ (where $\Lambda$ is the midpoint
along the $\Gamma$-{\rm K} path) in Fig.~\ref{fig:struct}(d). 
Similar to the case of
the NbSe$_{2}$ monolayers separated by vacuum \cite{SM},
$t_{\perp}^{\Gamma}$ $\gg$ $t_{\perp}^{\mathrm{K}}$.  With two monolayers
or more of CrBr$_3$ (as used by Kang {\it et al.} \cite{kang2021giant}),
$t_{\perp}$ at \textrm{K}
is suppressed significantly compared to $t_{\perp}$ at $\Gamma$ \cite{SM}.

Away from $\Gamma$, our calculations (Fig.~\ref{fig:struct}(d) show
that 2$t_{\perp}$ along the diagonal
($\Gamma$-{\rm K}), is lower compared to
2$t_{\perp}$ along the $\Gamma$-{\rm M} path.
Note that the magnitude of the spin-orbit coupling (SOC)
grows from $\Gamma$ to $\Lambda$,
while it is zero along $\Gamma$-{\rm M} \cite{ising_prx}.
Hence, the orbitals that contribute the least
to $2t_{\perp}$ leads to the largest $\Delta_{\rm SOC}$: the 
tunneling probability is correlated with the 
degree of the $z^2$ character of the Nb bands, while the SOC splitting is {\it anti}correlated
with it. This anticorrelation of the tunneling and SOC ``hot spots''
has crucial implications on interpreting tunneling 
measurements in these heterostructures, which we discuss next. 
Together with the spin filtering discussed above this constitutes another
theoretical prediction that is yet to be verified by experiment.

Armed with this quantitative understanding of $\lambda
_{\mathrm{ex}}$, $t_{\perp}^{\Gamma}$, and $t_{\perp}^{\mathrm{K}}$ we proceed to perform
model Hamiltonian calculations to describe tunneling across an Ising superconductor - ferromagnetic insulator - Ising superconductor 
junction \cite{SM}. 
We first consider, at a heuristic level, the impact of a magnetic exchange
field that is out-of-plane (parallel to $\hat{z}$)
and in-plane (parallel to $\hat{x}$) on the conductance peak.
For a magnetic exchange field, $B$, that is out-of-plane ($B \parallel c$)
the SOC and $B$ both
polarize the electron spins along $\hat{z}$
regardless of the in-plane momentum. Hence,
the magnetic exchange interaction reduces the energy of the singlet Cooper
pairs, while SOC plays no role. This leads to a familiar Pauli limited
superconductivity where the critical magnetic field is of the order of the
gap, $\Delta$ \cite{Maki1964}. Furthermore, the Cooper pairs retain their
singlet identity and are immune to the disorder scattering in accordance with
the Anderson theorem \cite{anderson1959theory}.

In contrast, when $B \perp c$,
the impact on the order parameter is weak due to the strong $\Delta
_{\mathrm{SOC}}$. 
However, the broadening of the conductance peak is sensitive to
$B \perp c$ and grows with the magnitude of
$B$.  When $B \perp c$, the spins 
acquire a finite in-plane momentum dependent component.
The spin
tilt angle is determined by 
$\Delta_{\mathrm{SOC}}$ and therefore varies, with the in-plane momentum.
Hence,
the paramagnetic defects behave as magnetic defects due to a finite in-plane
$B$ \cite{sosenko2017unconventional,Haim2020}.
These qualitative considerations are summarized in Table \ref{tab:summary}.

\begin{figure}[!th]
\includegraphics[width=8.5cm]{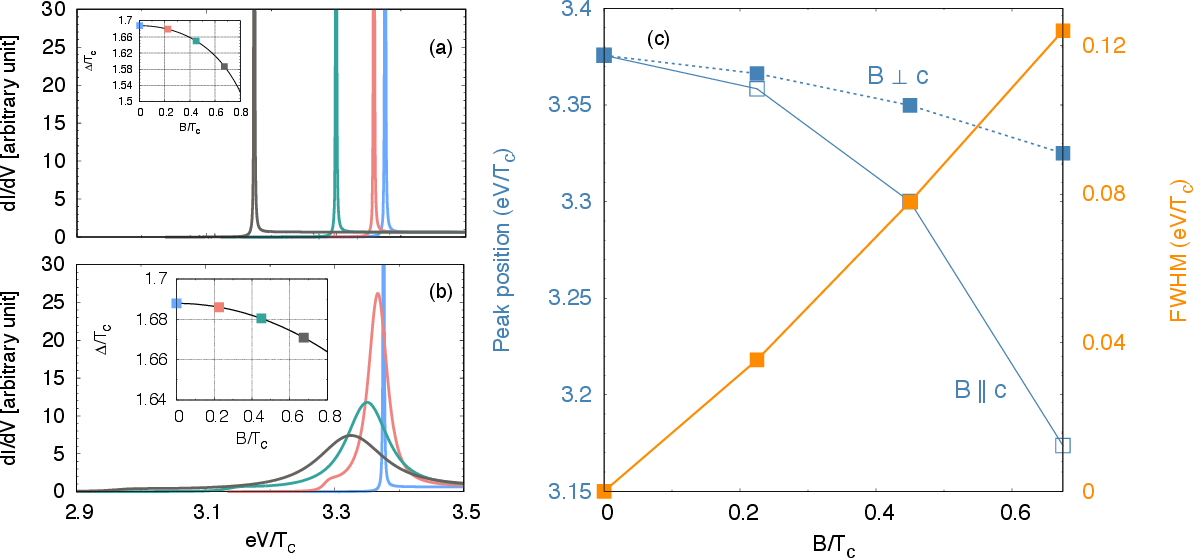}\caption{
Differential conductance $d
I/d V$ as a function of the bias voltage, $|e|V$ for an
(a) out-of-plane magnetic field that is along $\hat{z}$ ($B \parallel c$)
and an (b) in-plane magnetic field that is along $\hat{x}$ ($B \perp c$).  
We use four values of the
magnetic exchange field, $B$ that corresponds to
$B$ = 0 $T_{c}$ (blue), 0.225 $T_c$ (red), 0.45 $T_{c}$ (green), 
and 0.67 $T_{c}$ (grey) to determine the change in differential conductance
as a function of $B$. 
We use T=0.5$T_c$, $\Delta_{\rm SOC}$=20$T_c$ and a scattering rate, $\eta$=$T_c$
for all of our differential conductance calculations.
The inset in each panel (a) and (b) shows the suppression of the order
parameter $\Delta_{o}$ as a function of $B$.
(c)  Change in the peak position of the differential conductance
when $B \parallel c$ and $B \perp c$ (left vertical axis)
and the change
in the FWHM of the differential conductance as a function of
$B \perp c$ (right vertical axis).
}

\label{fig:cond}%
\end{figure}
\begin{table}[h]
\caption{Parameters that control the position and broadening of the
conductance peak as a function of the magnetic exchange field that is
out-of-plane, $B \parallel c$ (for moderately low temperatures) 
and in-plane, $B \perp c$. $B$ denotes the
magnitude of the exchange field, $\Delta_o$ is the order parameter,
$\Delta_{\mathrm{SOC}}$ is the magnitude of spin-orbit coupling, 
and $\eta \ll \Delta_{\mathrm{SOC}}$ is
the disorder scattering rate.}
\label{tab:summary}
\begin{ruledtabular}
\begin{tabular}{ccc}
& $B \parallel c$ & $B \perp c$    \\
\hline
peak shift & $(B/\Delta_o)^2$ & $(B/\Delta_{\mathrm{SOC}})^2$   \\
peak broadening & $0$ & $\eta B^2/(B^2 +  \Delta_{\mathrm{SOC}}^2)$  \\
\end{tabular}
\end{ruledtabular}
\end{table}

To put this on a firm theoretical
footing we use a model band dispersion to describe the
$\Gamma$-valley of monolayer NbSe$_{2}$ (since this is the valley 
through which most of the tunneling occurs!).
We use this to calculate the order parameter for $B \parallel c$ and
$B \perp c$ and combine this with the information
from our first-principles calculations to determine the spin-dependent tunneling
conductance \cite{SM}.

The calculated $d I/dV$ for an out-of-plane $B$ is shown in
Fig.~\ref{fig:cond}(a).
For each value of $B \parallel c$, we only find one $dI/dV$ peak at
$|e|V = 2 \Delta_{o}$ which is not split by Zeeman coupling. This is due to the fact
that the top and bottom NbSe$_{2}$ layers undergo the same amount of 
exchange splitting, $\lambda_{\mathrm{ex}}$ \cite{SM}. Hence, the superconducting density of states 
is split by the same amount and spin is conserved during tunneling
which leads to the single peak \cite{meservey1994spin}.

From Figure \ref{fig:cond}(a) it is evident that as the magnitude of
$B$ increases, the position of the 
$dI/dV$ peak decreases.
This is due to the suppression of the order parameter, $\Delta_{o}$, which is
proportional to $B^{2}$.  This is in contrast to the Zeeman split
peaks in the density of states which shifts linearly with the magnitude of the exchange field.
We also find that the full-width half maximum (FWHM) of the conductance peaks remains unchanged and is
insensitive to the amount of disorder that we consider.  

In Fig.~\ref{fig:cond}(b) we illustrate the calculated $d I/dV$ when $B \perp c$. 
We find a number of striking changes compared to Figure \ref{fig:cond}(a).
The peak position of the $d I/dV$ decreases and is weakly dependent on the magnitude
of $B$.
Secondly, the
FWHM of the $d I/dV$ increases as the magnitude of the in-plane
$B$ increases. This is consistent with the spin-flip scattering
rate increasing quadratically as $\eta B^{2}/2\Delta_{\mathrm{SOC}}^{2} $, where
$\eta$ is the scattering rate due to paramagnetic defects
\cite{sosenko2017unconventional,Haim2020}.

In Figure \ref{fig:cond}(c) we summarize our calculations of the peak position
and FWHM as a function of the magnitude and direction of $B$. 
These results confirm the
qualitative analysis in Table \ref{tab:summary} and provide a physically intuitive explanation for the modest 
increase in $\Delta$, accompanied by the coherence peak broadening observed in
tunneling measurements \cite{kang2021giant}.

\begin{figure}[h]
\includegraphics[width=8.5cm]{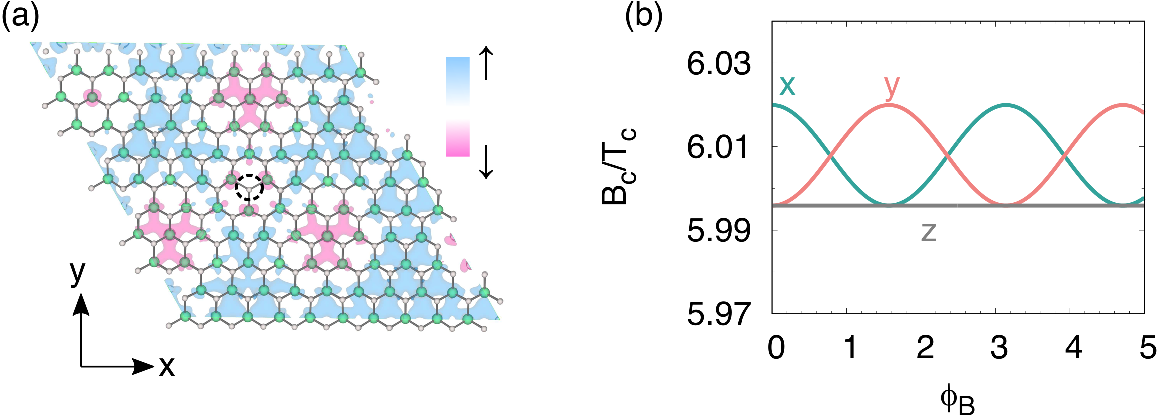}\caption{
(a) Spin density of a single selenium
vacancy within a 10$\times$10$\times$1 supercell of monolayer NbSe$_{2}$. The
different colors correspond to different signs of the magnetization. The net
magnetization is $\sim$ 0.6 $\mu_{B}$.
The position of the missing selenium atom is denoted with the black dotted circle. 
 (b) The in-plane critical field,
$B_{c}$ as a function of the field orientation, specified by the angle
$\phi_{B}$ formed by the magnetic field with respect to $x$-direction. We
consider the magnetic easy axis of the defect spin along $\hat{x}$ (green),
$\hat{y}$ (red) and $\hat{z}$ (grey) with spin-flip scattering rates $\eta
_{1}$,$\eta_{2}$ and $\eta_{3}$ equal to 0.25 T$_{c}$. We set T = 0.2T$_{c}$
and $\Delta_{\mathrm{SOC}}$ = 20 T$_{c}$. 
}
\label{fig:two-fold}%
\end{figure}

We now include two additional effects that are likely present in
NbSe$_{2}$: magnetic point defects and extended defects. 
One candidate for magnetic
point defects are Se
vacancies, $V_{\rm Se}$, which have been found in appreciable concentrations in 
NbSe$_{2}$ \cite{nguyen2017atomic}. 
To verify this hypothesis we performed spin-unrestricted
first-principles calculations of $V_{\rm Se}$ in a (10$\times$10$\times$1) supercell.  
We find a sizeable magnetization ($\approx
0.6\ \mu_{B}$, within our 300 atom supercell) and the induced
magnetization has a finite length scale that is commensurate with the in-plane lattice
constant ($\sim$ 15 \AA) of
our large supercell as illustrated in Fig.~\ref{fig:two-fold}(a).  This is likely due to 
monolayer NbSe$_2$ being close to a magnetic instability \cite{ising_prx,divilov2021magnetic}.
Interestingly, the induced spin-polarization is large and sign-changing, reminiscent of
Friedel oscillations.

While we did not compute the magnetic anisotropy of such defects, it is likely to be easy-axis.  
Indeed,
for an ideal hexagonal lattice, the symmetry allows for mixing of the 
$x^{2}-y^{2}$ and $xy$ Nb $d$-orbitals.  This mixing can generate an orbital moment
$L_{z},$ with no cost in kinetic energy. Hence, an isolated $V_{\rm Se}$ defect is likely to
have its magnetic moment oriented along $\hat{z}$.

If the magnetic moment of the defect remains
along $\hat{z}$, it has a pair breaking effect in the same way as it would in 
an ordinary $s$-wave superconductor.  However, if it
is aligned in-plane, this pair-breaking effect, within the Born approximation, 
becomes strongly anisotropic, leading to a 
considerable enhancement of the in-plane critical field in the in-plane direction parallel to the impurity moment.
\cite{mockli_JAP20}. In Fig.~\ref{fig:two-fold}(b) we illustrate the in-plane critical
field as a function of the orientation of $B\perp c$. Note the two-fold oscillations 
for an in-plane defect spin.

The finite spatial extent of the magnetization, $R_{d}$, due to $V_{\rm Se}$, also
provides a plausible explanation for the 
puzzling hysteresis in the tunneling conductance that occurs at
$T\lesssim(T_{c}-2$K$)$ \cite{kang2021giant}.
As the temperature is lowered below $T_{c}$, the superconducting
coherence length, $\xi$, decreases and at some point may become lower than $R_{d}$. 
When $\xi<R_{d}$, scattering would occur within the
unitary limit which would result in superconductivity being suppressed near
the vacancy, within a length of the order of $R_{d}$. This suppression only
occurs when the magnetic moment of the defect is oriented along $\hat{z}$ (as
discussed above, this is likely the case for isolated $V_{\rm Se}$). When the
pairing energy of the resulting \textquotedblleft puddle\textquotedblright\ of
finite magnetization, which is $\sim\Delta^{2}N(0)R_{d}^{2}$, becomes larger
than the magnetic anisotropy energy (typically on the order of $\mu$eV for
point defects), the magnetic moment of the point defect would flop to be
in-plane. We expect this behavior to be hysteretic, as is typical for a magnetic transition.

So far we have considered point defects. However, as-grown NbSe$_{2}$ is
known to exhibit extended defects such as grain boundaries \cite{wang2017high} and dislocations.
Elastic fields tend to align linear defects along the same direction, which would break the
global $C_{3}$ symmetry of the hexagonal lattice. 
The strain fields that manifest from extended defects have been proposed
to affect the symmetry of the superconducting state \cite{willa2020inhomogeneous}.
Below we present an alternative mechanism on how extended defects might
break the $C_{3}$ symmetry in the superconducting state.

At first glance it seems that this would require a ``nematic'' superconducting order
parameter, that intrinsically breaks the $C_{3}$ symmetry \cite{hamill2021two,cho2020distinct} .
While this would, by definition, generate the desired symmetry breaking, it also implies 
that the expected $s$-wave state is nearly degenerate with some other state(s) with 
a different pairing symmetry. This is a logical assumption in materials 
like the Fe-based superconductors, where the same spin-fluctuations generate pairing in the $s_\pm$ 
and a $d$ channel, so it is not surprising that a combination of both may be energetically favorable.
In the superconducting TMDs, on the other hand,
spin or Coulomb interactions are pair-breaking in the $s$-wave channel.
As a result, the competition
between conventional and unconventional paring mechanisms requires an extremely fine tuning of
parameters, and a dramatic difference between the bulk and the single layer pairing mechanism.
In addition, one must assume that the interactions
are extremely sensitive to the small strain generated by the
extended defects or externally.  

In this context, an interesting question to ask is: can
symmetry-breaking extended defects result in a tunneling
conductance
and critical field that has $C_{2}$ symmetry with respect to the direction of
the external magnetic field \textit{without impacting the symmetry of the
superconducting order parameter?}
To this effect, we observe that while an isolated point defect 
(vacancy) is expected to have its spin aligned with the $z$-axis,
the same does not hold  near an extended defect, where the 
local $C_{3}$ symmetry is broken and the $d$-orbitals of the Nb dangling
bond states can mix.  In this case the
orbital magnetic moment can point along an in-plane direction determined by the
linear defect.

According to our theory, the defect-induced broadening of the
tunneling peaks
and the pair breaking by the magnetic field will depend on the angle between
the direction of the applied magnetic field and the orientation of the
extended defect within the basal plane of NbSe$_{2}$.
The extended defects broaden the superconducting density of states near the
conductance peak, break the $C_{3}$ rotational symmetry at or slightly below
$T_{c}$, in agreement with the existing experimental observations
\cite{cho2020distinct,hamill2021two}.  If these point defects have a finite
magnetic moment, they can also indirectly
trigger an anisotropy in the magnetoresistance near $T_{c}$ by generating an
easy-axis magnetic anisotropy of the defect.

This provides an immediate explanation of the $\pi$-periodic angular
dependence (\textit{i.e., a} $C_{2}$ rather than $C_{3}$ symmetry) the in-plane magnetoresistance either in
the transition region centered at $T_{c}$ \cite{hamill2021two} or
slightly below $T_{c}$ \cite{cho2020distinct}, without invoking an {\it
ad hoc} assumption about nematic superconductivity (admittedly, our interpretation
assumes an in-plane easy axis for magnetic defects 
pinned to extended defects, but this is plausible from a 
materials science point of view.)

In summary, using first-principles calculations and an analytical theory for Ising superconductivity
we have systematically investigated proximity induced effects in 
NbSe$_2$/CrBr$_3$ heterostructures.  We find
CrBr$_{3}$ leads to a proximity-induced exchange 
splitting of the NbSe$_{2}$ states and that the NbSe$_2$ states at $\Gamma$
contribute the most to tunneling.
Scattering of the NbSe$_2$ states at $\Gamma$ off of paramagnetic point defects
leads to a pronounced
broadening of the tunneling peaks, a modest enhancement of the
superconducting gap when the magnetic exchange field is in-plane.
Within the same framework, extended linear defects generate
two-fold oscillations of the critical field, seen in experiments. 
Finally, we find point defects such as selenium vacancies acquire a finite magnetization
of a sizeable length scale, which can explain the finite hysteresis 
in the conductance. Last but not least, we predict two effects that 
can be verified by future experiments (and may have more theoretical ramifications than
discussed here): spin-filtering when tunneling through the CrBr$_3$ barrier and anticorrelation 
between the SOC and the tunneling probability.

\begin{acknowledgements}
We thank David M\"ockli, Jie Shan, Kin Fai Mak and Rafael Fernandes for helpful discussions.
D.W was supported by the Office of Naval Research through the Naval Research
Laboratory's Basic Research Program. I.I.M. was supported by ONR through grant
N00014-20-1-2345. M.H. and M.K. acknowledge the financial support from the
Israel Science Foundation, Grant No. 2665/20. Calculations by D.W. and
I.I.M. were performed at the DoD Major Shared Resource
Center at AFRL.
\end{acknowledgements}

%\nocite{Blochl_PAW}
%\nocite{VASP_Ref}
%\nocite{VASP_ref2}
%\nocite{perdew1996generalized}
%\nocite{grimme2010consistent}
%\nocite{grimme2006semiempirical}
%\nocite{grimme2011effect}
%\nocite{klimevs2011van}
%\nocite{handy1952structural}
%\nocite{mazin2001tunneling}
%\nocite{Haim2020}
%\bibliography{BIBLIO}


\end{document}